\documentclass[prd,nofootinbib]{revtex4}

\usepackage{amsmath,amssymb}
\usepackage{mathrsfs}
\usepackage{pstricks}
\usepackage{pst-func}
\usepackage{graphicx}
\usepackage{refcount}
\usepackage{cases}
\usepackage{slashed}

\begin{document}

\title{Comment on `Relativistic wave-particle duality for spinors'}

\author{Carl F. Diether, III}

\email{fred.diether@einstein-physics.org}
 
\author{Joy Christian}

\email{jjc@bu.edu}

\affiliation{Einstein Centre for Local-Realistic Physics, 15 Thackley End, Oxford OX2 6LB, United Kingdom}


\begin{abstract}
In a recent paper \cite{Poplawski}, it has been proposed that relativistic wave–particle duality can be embodied in a relation that shows that the four-velocity of a particle is proportional to the Dirac four-current. In this note we bring out some problems with that idea. In particular, we point out that, in line with existing literature on Einstein-Cartan gravity with torsion, the spin-torsion term should represent negative energy. Moreover, if what is proposed in Eq.~(20) of \cite{Poplawski} is correct, then the spin torsion term would be zero in the rest frame of the fermion.  We suggest a possible resolution of that dilemma.
\end{abstract}

\maketitle

\parskip 6pt
\baselineskip 12pt

In a recent paper \cite{Poplawski}, Poplawski has proposed that relativistic wave–particle duality can be embodied in a relation that shows that the four-velocity of a particle is proportional to the Dirac four-current:
\begin{equation}
\langle u^i\rangle=\frac{\bar{\psi}\gamma^i\psi}{\bar{\psi}\psi}. \label{velocity}
\end{equation}
While this is an interesting proposal, it is puzzling why the non-linear term associated with spin-torsion in Eqs.~(12) of \cite{Poplawski} is positive when elsewhere in the literature it appears negative in the same context, for example in \cite{Diether,diether2,Hehl-Datta,Freidel,Perez,Freidel-2}. With $\hbar = c = 1$ and $\kappa = 8\pi G$, here are the equations in question, known as the Hehl-Datta equations:
\begin{equation}
i\gamma^k \psi_{:k} + \frac{3\kappa}{8}(\bar{\psi}\gamma_k \gamma^5 \psi)\gamma^k \gamma^5 \psi = m\psi,\quad -i \bar{\psi}_{:k} \gamma^k + \frac{3\kappa}{8}(\bar{\psi}\gamma_k \gamma^5 \psi)\bar{\psi}\gamma^k \gamma^5  = m\bar{\psi}. \label{pos}
\end{equation}
By contrast, with negative non-linear term in the above equations, quantum electrodynamics can be completed in a natural way and predicts the ``size" of elementary fermions to be near the Planck length, as we have shown in \cite{Diether,diether2}. Moreover, rather remarkably, in that case no regularization or re-normalization is required for certain solutions. But the positive non-linear term in (\ref{pos}) exasperates the so-called infinite self-energy of an electron instead of solving it. 

To understand the difference better, let us begin with the pioneering derivation by Hehl and Datta of the first of the above equations in \cite{Hehl-Datta}, namely, their equation (C'), which reads as follows,
\begin{equation}
    \left[\gamma^{\alpha} \nabla^{\{\}}_{\alpha}- \frac{3}{8}i l^2(\psi^{+}\gamma_5 \gamma^{\alpha}\psi)\gamma_5 \gamma_{\alpha}\right]\psi = i m\psi, \label{hd}
\end{equation}
where $l$ is Planck length. However, when we multiply through this equation by $-i$ we obtain
\begin{equation}
    \left[- i\gamma^{\alpha} \nabla^{\{\}}_{\alpha}- \frac{3}{8} l^2(\psi^{+}\gamma_5 \gamma^{\alpha}\psi)\gamma_5 \gamma_{\alpha}\right]\psi = m\psi\,.
\end{equation}
The negative sign on the kinetic energy term is due to the Lagrangian density that they started with in \cite{Hehl-Datta}. If one starts with a different Lagrangian density, then the result, in Poplawski's notation, can be
\begin{equation}
i\gamma^{\alpha} \nabla_{:\alpha}\psi - \frac{3}{8} l^2(\bar{\psi}\gamma_5 \gamma^{\alpha}\psi)\gamma_5 \gamma_{\alpha}\psi = m\psi\,,
\end{equation}
where the colon denotes the covariant derivative.  Which we believe is the proper form for the equation and it also agrees with the action Eq.~(2.1) of this paper \cite{Freidel}. However, the sign on the derivative term does not matter because it vanishes for the rest frame analysis we will be concerned with, as in \cite{Diether}.

Another example where negative non-linear term is used is a paper by Perez and Rovelli \cite{Perez}. Their Eq.~(15) reads:
\begin{equation}
S_{int}[e, \psi] = -\frac{3}{2}\pi G \frac{\gamma^2}{\gamma^2+1}\int{d^4x\, e\, (\bar{\psi}\gamma_5\gamma_A \psi)(\bar{\psi}\gamma_5\gamma^A \psi)}.
\end{equation}
The Immirzi parameter, $\gamma$, can be taken to infinity, and, with $\kappa = 8\pi G$, the standard minimal coupling is recovered:
\begin{equation}
    S_{int}[e, \psi] = -\frac{3\kappa}{16} \int{d^4x\, e\, (\bar{\psi}\gamma_5\gamma_A \psi)(\bar{\psi}\gamma_5\gamma^A \psi)}.
\end{equation}
In our view, this is the proper form for the action, which makes it clear that the term in question is negative in the action. The authors seem to have derived this term differently, but, unfortunately, they do not provide the details.

According to the above analysis, Poplawski in \cite{Poplawski} has used the wrong sign in the non-linear term in its Eq.~(12). 

Another issue concerns Eq.~(20) in \cite{Poplawski} where it is shown, for the rest frame, that the middle term in the non-linear term $(\bar{\psi}\gamma_k\gamma5\psi)$ vanishes, $(\bar{\psi}\gamma_0\gamma5\psi) = 0$, which makes the whole term vanish in the rest frame. By contrast, we have found that fermion anti-fermion mixing is involved in the non-linear term in the rest frame of the particle, as we demonstrate below in the appendix. Moreover, in the full machinery of quantum field theory, $\psi$ can represent particle or anti-particle in the rest frame, as in the following equations:
\begin{equation}
\psi(0, t)\mid i\,\rangle = \sqrt{\frac{1}{r^3}} \,u^i(m)e^{-iEt} \quad \text{or} \quad \psi(0, t)\mid i\,\rangle = \sqrt{\frac{1}{r^3}} \,v^i(m)e^{+iEt},
\end{equation}
where the $\mid i\,\rangle$ represents the initial state. In order for the non-linear term to be non-zero in the rest frame, one must have one of the $\psi$'s to represent an appropriate anti-fermion, as in the following equations: 
\begin{equation}
\left(\begin{array}{cccc}
1& 0& 0& 0
\end{array}\right)
\gamma^5 \left(\begin{array}{c}
0\\
0\\
1\\
0 
\end{array}\right) = 1\quad \text{or} \quad
\left(\begin{array}{cccc}
0& 0& 1& 0
\end{array}\right)
\gamma^5 \left(\begin{array}{c}
1 \\
0 \\
0 \\
0 
\end{array}\right) = 1.
\end{equation}
Consequently, Eqs.~(21) and (22) in \cite{Poplawski} also appear to be incorrect.

\subsection*{Note added to proof (08/10/2021):}

In a recent paper \cite{Poplawski2}, Grosso and Poplawski have proposed that gravitational torsion can regularize the self-energy of charged leptons and give the bare masses of them.  We also tried to use the ``standard'' text book electron self-energy\break originally proposed by Weisskopf in 1939 to solve for bare masses using gravitational torsion via the Hehl-Datta equation (\ref{hd}).  But after a few attempts, we realized it was completely the wrong process to use.  The proper process is to analyze in the rest frame only of the charged fermion.  The clue came from our semi-classical analysis which can be found in the appendix below. That means that there is no linear momentum at all and there are no propagators to use from Feynman diagrams.  According to our process, there is no need to have to regularize the self-energy \cite{Diether} since the spin-torsion potential is completely cancelled out by the electromagnetic potential, leaving the rest mass only.  So, in our view, Grosso and Poplawski are on the wrong track.  There are no ``bare'' masses for the charged fermions.

Furthermore, we believe that we have found the source of Poplawski's sign flip for the spin-torsion term.  It is in Eq.~(2.6.55) of his book \cite{Poplawski3}.  In the last term of the first line of that equation, an imaginary, ``i", appears for no reason.

\section*{Acknowledgement}
The authors would like to acknowledge that it was Nikodem Poplawski's earlier research in Einstein-Cartan theory of gravity that motivated us to investigate the Dirac-Hehl-Datta equation more deeply. We would also like to thank Luca Fabbri for discussion about the sign issues raised in the previous version of this paper. 

\appendix*
\section{Semi-Classical Derivation of Charged Fermion Self-Energy}
In the rest frame, where normal gravity is effectively zero, with natural units $\hbar = c = 1$, along with the electrostatic term via a local gauge transformation, we have
\begin{equation}
i \,\gamma^0 \frac{\partial\psi}{\partial t\,} + q A_0\,\gamma^0\psi = m \, \psi + \frac{3\kappa}{8} \left(\bar{\psi}\gamma^5 \gamma_0\psi\right) \gamma^5\gamma^0\psi\,, \label{restbc}
\end{equation}
which can be further simplified to
\begin{equation}
i\left(\begin{array}{c}
+\frac{\partial\psi_1}{\partial t\;} \\ \\
+\frac{\partial\psi_2}{\partial t\;} \\ \\
-\frac{\partial \psi_3}{\partial t\;} \\ \\
-\frac{\partial \psi_4}{\partial t\;} 
\end{array}\right)
+ q A_0
\left(\begin{array}{c}
+\psi_1 \\ \\
+\psi_2 \\ \\
-\psi_3 \\ \\
-\psi_4
\end{array}\right)
= m
\left(\begin{array}{c}
+\psi_1 \\ \\
+\psi_2 \\ \\
+\psi_3 \\ \\
+\psi_4
\end{array}\right)
- \frac{3\kappa}{8}\left\{\psi_1^*\psi_3+\psi_2^*\psi_4+\psi_1\psi_3^*+\psi_2\psi_4^* \right\}
\left(\begin{array}{c}
-\psi_3 \\ \\
-\psi_4 \\ \\
+\psi_1 \\ \\
+\psi_2
\end{array}\right), \label{A2}
\end{equation}
where we have used 
\begin{equation}
\gamma^0 = \left(\begin{array}{cccc}
+1& 0& 0& 0\\
0& +1& 0& 0\\
0& 0& -1& 0\\
0& 0& 0& -1
\end{array}\right)\;\;\;\;\;\;\;\;\;\;\text{and}\;\;\;\;\;\;\;\;
\gamma^5= \left(\begin{array}{cccc}
0& 0& +1& 0 \\
0& 0& 0& +1 \\
+1& 0& 0& 0 \\
0& +1& 0& 0
\end{array}\right).
\end{equation}
If we now represent the particles and anti-particles with two-component spinors ${\psi_a}$ and ${\psi_b}$, respectively \cite{Griffiths}, where
\begin{equation}
\psi_a := \left(\begin{array}{c}
\psi_1 \\
\psi_2 \end{array}\right)
\;\;\;\text{and}\;\;\;
\psi_b := \left(\begin{array}{c}
\psi_3 \\
\psi_4 \end{array}\right) \label{c444}
\end{equation}
are the two-component spinors constituting the four-component Dirac spinor, then the above equation can be written as two {\it coupled} partial differential equations:
\begin{align}
+\,i \,\frac{\partial \psi_a}{\partial t\;} + q A_0 \,\psi_a &= m\,\psi_a + \frac{3\kappa}{8} \left\{\psi_1^*\psi_3+\psi_2^*\psi_4+\psi_1\psi_3^*+\psi_2\psi_4^*\right\}\,\psi_b \label{10bc}\\
-\,i \,\frac{\partial \psi_b}{\partial t\;} - q A_0 \,\psi_b &= m\,\psi_b - \frac{3\kappa}{8} \left\{\psi_1^*\psi_3+\psi_2^*\psi_4+\psi_1\psi_3^*+\psi_2\psi_4^*\right\}\,\psi_a\,. \label{11bc}
\end{align}
It is now very easy to see the fermion anti-fermion mixing in the non-linear term, which can be seen also in Eq.~(\ref{A2}).

Unlike the case in Dirac equation, these equations for the spinors ${\psi_a}$ and ${\psi_b}$ are coupled equations even in the rest frame. They decouple in the limit when the torsion-induced axial-axial self-interaction is negligible. On the other hand, at low energies it is reasonable to assume that, in analogy with the Dirac spinors in flat spacetime, the above two-component spinors for free particles decouple in the rest frame, admitting plane wave solutions of the form
\begin{equation}
\psi_a(t) = \sqrt{\frac{1}{V}}\,e^{-iEt} \psi_a(0) \;\;\;\;\;\text{and}\;\;\;\;\;
\psi_b(t) = \sqrt{\frac{1}{V}}\,e^{+iEt} \psi_b(0), \label{psiabc}
\end{equation}
where $E=m=\omega = 2\pi/t$ in the rest frame.  We now note that in the rest frame the derivative term in the eqs. (\ref{10bc}) and (\ref{11bc}) vanishes since $\psi$ is constant and the kinetic energy in the rest frame would be zeo anyways, also because in the rest frame the probability of finding the particle in a given volume at time ${t}$ is one. Moreover $E = 2\pi/t$ gives
\begin{equation}
\psi_a(t) = \psi_b(t). \label{14bc}
\end{equation}
Consequently eqs. (\ref{10bc}) and (\ref{11bc}) uncouple and simplify to 
\begin{align}
+ q A_0 \,\psi_a &= m\,\psi_a + \frac{3\kappa}{8} \left\{\psi_1^*\psi_1+\psi_2^*\psi_2+\psi_3\psi_3^*+\psi_4\psi_4^*\right\}\,\psi_a\,, \label{10bcu}\\
- q A_0 \,\psi_b &= m\,\psi_b - \frac{3\kappa}{8} \left\{\psi_1^*\psi_1+\psi_2^*\psi_2+\psi_3\psi_3^*+\psi_4\psi_4^*\right\}\,\psi_b\,. \label{11bcu}
\end{align}
Substituting the $\psi$'s from eq. (\ref{psiabc}) and simplifying reduces eqs. (\ref{10bcu}) and (\ref{11bcu}) to the following pair of equations:
\begin{align}
+ q A_0 \psi_a(0) &= m\,\psi_a(0) + \frac{3\kappa}{8\,r^3}|\psi(0)|^2 \,\psi_a(0) \label{16abc} \\
\text{and}\;\;\;
- q A_0 \psi_b(0) &= m\,\psi_b(0) - \frac{3\kappa}{8\,r^3}\, |\psi(0)|^2 \,\psi_b(0)\,. \label{16bbc}
\end{align}
And since $|\psi(0)|^2 = 1$ in the rest frame, these equations can be further simplified to
\begin{align}
+\,q A_0 &= m + \frac{3\kappa}{8\,r^3}\,,  \\
\text{and}\;\;\;
-\,q A_0 &= m - \frac{3\kappa}{8\,r^3}\,.
\end{align}
Substituting in natural units for the scalar field ${A_0= V = q /(4\pi r)}$ in the Lorentz gauge (where ${V}$ is the electric potential), and for $\kappa= 8\pi G$, we finally arrive at our central equations, for any electroweak fermion of charge ${q}$ and mass ${m}$ and its anti-particle in the Riemann-Cartan spacetime:
\begin{align}
+\frac{q^2}{4\pi r} - \frac{3\pi G}{r^3} &= +m  \label{resultbc} \\
\text{and}\;\;\;\;
-\frac{q^2}{4\pi r} + \frac{3\pi G}{r^3} &= +m, \label{anti-resultbc}
\end{align}
where ${r}$ is the radial distance from ${q}$ and the two equations correspond to the particle and anti-particle, respectively.  Finally, replacing $q$ with electron charge and replacing $\hbar$ and $c$ gives us our central equation for a fermion for our semi-classical evaluation:
\begin{equation}
\frac{\alpha \hbar c}{r} - \frac{3\kappa (\hbar c)^2}{8\,r^3} = m c^2.  \label{result2bc} 
\end{equation}
This same equation is also derived via a S-matrix quantum field theory evaluation for the rest frame in \cite{Diether}.

\end{document}